# Understanding High-Field Electron Transport Properties of Monolayer Transition Metal Dichalcogenides and Strain Effects

Chenmu Zhang, Long Cheng, Yuanyue Liu*

Department of Mechanical Engineering and Texas Materials Institute

The University of Texas at Austin, Austin, TX, 78712

*yuanyue.liu@austin.utexas.edu

**Abstract:**
Monolayer transition metal dichalcogenides ($MX_2$) are promising candidates for future electronics. Although the transport properties (e.g. mobility) at low electric field have been widely studied, there are limited studies on high-field properties, which are important for many applications. Particularly, there is lack of understanding of the physical origins underlying the property differences across different $MX_2$. Here by combining first-principles calculations with Monte Carlo simulations, we study the high-field electron transport in defects-free unstrained and tensilely strained $MX_2$ (M=Mo, W and X=S, Se). We find that $WS_2$ has the highest peak velocity (due to its smallest effective mass) that can be reached at the lowest electric field (owing to its highest mobility). Strain can increase the peak velocity by increasing the scattering energy. After reaching the peak velocity, most $MX_2$ demonstrates negative differential mobility (NDM). $WS_2$ shows the largest NDM among unstrained $MX_2$ due to the strongest effect of electron transfer from the low-energy small-mass valley to the high-energy large-mass valley. The tensile strain increases the valley separation, which on one hand suppresses the electron transfer in $WS_2$, on the other hand allows the electrons to access the non-parabolic band region of the low-energy valley. The latter effect leads to an NDM for electrons in the low-energy valley, which can significantly increase the overall NDM at moderate strain. The valley-separation induced NDM in the low-energy valley is found to be a general phenomenon. Our work unveils the physical factors underlying the differences in high-field transport properties of different $MX_2$, and also identifies the most promising candidate as well as effective approach for further improvement.

**Introduction:**

Monolayer (2D) semiconducting transition metal dichalcogenides ($MX_2$) have gained extensive interest for electronics due to their promising properties, such as gate-tunability, [1,2] mechanical flexibility, [3,4] and the ease for being assembled for heterostructures. [5] Although the transport properties (e.g. mobility) at low electric field ($E$) have been widely studied, [6–9] there are limited studies on high-field properties, [10–12] which are important for many applications. For example, the maximum intrinsic frequency of many electronic devices is closely related with the maximum average velocity of the charge carriers, [13] which is reached at high $E$. Moreover, electronic oscillators and amplifiers sometimes utilize the negative differential mobility (NDM) property of semiconductors at high $E$. In semiconductors with multiple energetically close valleys in the electronic structures (e.g., GaAs), the high $E$ can transfer carriers from the lower-energy and lower-mass (higher-velocity) valley, leading to the NDM in the bulk velocity-field curve. This intervalley transfer of electrons (sometimes referred to as the Gunn effect) is the basis of operation of the Gunn diode, whose current-voltage characteristic displays a region of negative differential resistance (NDR), wherein an increase in voltage results in a decrease in electrical current. These applications urge for an improved understanding of the high-$E$ carrier transport properties in $MX_2$, particularly, what causes the different properties across different $MX_2$.

The electron transport in various 2D materials and devices have been widely studied using different theoretical/computational methods, including Monte Carlo (MC), Wigner equation, nonequilibrium Green's functions, and master equation for the density matrix [14–16]. Particularly, the MC method has been applied to study silicene and germanene, [17] InSe, [18] and graphene. [19–23] Regarding the high-field electron transport in $MX_2$, Ferry [24] studied $MoS_2$ and $WS_2$ using MC method: the electron band structure and the scattering rates approximated by analytic models with parameters taken from first-principles calculations and experiments; $MoS_2$ is found to have a higher saturation velocity than $WS_2$. This contradicts the study by Kim et al, [12] which finds that the $WS_2$ has the highest saturation velocity compared with $MoS_2$, $WSe_2$ and $MoSe_2$, using full-band MC. However, Kim's work does not report NDM of $WS_2$, which disagrees with the experimental observation, [11] possibly because the $E$ range explored in this work is too small. Moreover, the number of points sampling the Brillouin zone is very small, while it has been shown that much finer grid is required to converge the results. [25] Therefore, to improve the understanding of the high-$E$ electron transport in $MX_2$ and the origins behind their differences, it is necessary to use more accurate methods to systematically calculate and analyze their properties.

In this work, we investigate the high-field electron transport properties of defects-free unstrained (denoted as 'us-') and strained (2% isotropic tensile strain; denoted as 's-') monolayer $MoS_2$, $MoSe_2$, $WS_2$, and $WSe_2$, by combining first-principles calculations and Monte Carlo (MC) simulations. We reveal the underlying physical factors that govern the properties in different $MX_2$, and identify the $WS_2$ as the best candidate for high-$E$ applications. Moreover, we find that the strain can increase the peak velocity and the NDM, and the origin of the strain effects is also unveiled. Particularly, the tensile strain increases the valley separation, which exposes the non-parabolic band region of the low-energy valley, and leads to an NDM of electrons in the low-energy valley. The valley-separation induced NDM in the low-energy valley is found to be a general phenomenon, suggesting a new design consideration for NDM.

**Methods:**
The MC method is a statistical method used to yield numerical solution to the Boltzmann transport equation (BTE) [26–32] which includes complex band structure and scattering processes. In contrast to low-field case where analytic solutions can be derived under certain approximations, in high field, the numerical method becomes necessary due to the nonlinear terms in BTE. In MC method, the carrier drifts from one state to another driven by the external electric field, and then is scattered stochastically to another state. This process repeats many times, generating a steady state of carrier distribution. [26] The drift motion in the momentum-space is described by:

$$q\mathbf{E}\tau = \hbar(\mathbf{k}_f - \mathbf{k}_i),\qquad(1)$$

where $\mathbf{k}_i$ and $\mathbf{k}_f$ are the wavevectors of the initial and final electronic states of drift respectively, and $\tau$ is the relaxation time that is determined stochastically from the transition rate of each electronic state along the drift path. The transition rate is the rate of transition from one electronic state to another, and an electronic state can be transited to many different states with different rates. In order to reduce the high computational demand of scattering process, Cellular Monte Carlo (CMC) [28,29,31] algorithm has been applied. Since the materials considered here are defects free, the scatterings are dominated by phonons and thus here we consider only phonon-assisted transitions. Most of previous studies [26–29,31,32] used analytic expressions (such as deformation-potential theory) derived from simplified models to approximate the transition rates. Here we directly calculate the transition rates from first principles. Specifically, the electron-phonon coupling (EPC) strengths ($g$) are calculated using Density Functional Perturbation Theory (DFPT) on a sparse k and q grid (9×9), [33] and then interpolated to a dense k and q grid (180×180). [34] Note that here the Fröhlich interaction is treated using 2D model, [35,36] different from the commonly used 3D model. [37] The transition rates are then obtained as:

$$P_{\mathbf{k},\mathbf{k}+\mathbf{q}} = \frac{2\pi}{\hbar}\sum_{j}|g_{\mathbf{k},\mathbf{k}+\mathbf{q},j}|^2\left[n_{\mathbf{q},j}\delta(\varepsilon_{\mathbf{k}+\mathbf{q}} - \varepsilon_{\mathbf{k}} - \hbar\omega_{\mathbf{q},j}) + (n_{\mathbf{q},j} + 1)\delta(\varepsilon_{\mathbf{k}+\mathbf{q}} - \varepsilon_{\mathbf{k}} + \hbar\omega_{\mathbf{q},j})\right],\qquad(2)$$

where $\mathbf{k}$ and $\mathbf{k}+\mathbf{q}$ are initial and final electron states, and $\mathbf{q}$ is the absorbed/emitted phonon wavevector, respectively. j is the index of phonon branch, $g_{\mathbf{k},\mathbf{k}+\mathbf{q},j}$ is the EPC strength and $n_{\mathbf{q},j}$ is the Bose-Einstein distribution. This approach is called full-band full-scattering approach, as both the electronic structure and the transition rates are fully determined from first principles. The full-band full-scattering approach is thus more accurate than the conventional MC approaches that assume analytical expressions for transition rates and/or band structure. [30] The full-band full-scattering approach has been applied to study $MX_2$, [12] and other 2D materials such as silicene, [17] germanene [17] and InSe. [18] Compared with those cases, here we use a much finer grid, which is necessary to converge the results [25] (see Fig. S9 for convergence test), meanwhile some numerical issues have to be addressed for computation efficiency. These details can be found in the Supplemental Material [38] (see, also, references [25,26,28,29,31,34,35,37,39-55] therein).

**Results and discussion:**

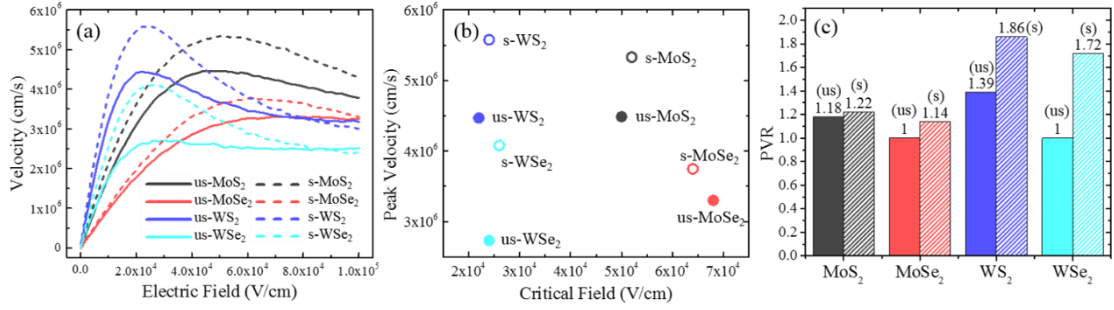

**Figure 1.** (a) Average velocity of electrons in unstrained (denoted as 'us-') and strained (denoted as 's-') MoS$_2$, MoSe$_2$, WS$_2$ and WSe$_2$ monolayers. (b) Peak velocity and critical field for each system. (c) Peak-to-valley ratio (PVR) for each system. A unity PVR means no negative differential mobility.

Fig. 1a shows the average velocity $\langle \mathbf{v} \rangle$ of electrons as a function of the electric field, for both unstrained and strained MX$_2$. The $\langle ... \rangle$ indicates an average over electronic states at steady state, and $\mathbf{v} = (d\varepsilon/d\mathbf{k})/\hbar$. Here we choose the electric field to be along the zigzag (x) direction of MX$_2$, and record the velocity component along the x direction. We limit the $E$ to be $<10^5$ V/cm as a higher field may break the material. [10,11] At low $E$, the $\langle \mathbf{v} \rangle$ increases linearly with $E$, the slope of which defines the mobility. The extracted mobilities agree with those calculated by iteratively solving the BTE [25] under the low-field limit approximation (Fig. S5), validating our MC method. The mobility follows the order: us-WS$_2$>us-MoS$_2$>us-WSe$_2$>us-MoSe$_2$, and the tensile strain increases the mobility for all the materials. The trend of the mobility has been explained in early work, [9] which highlights the role of Born charge (which determines the scattering by longitudinal optical phonons). The $\langle \mathbf{v} \rangle$ reaches a peak value $\langle \mathbf{v} \rangle_p$ at a certain field $E_c$. The $\langle \mathbf{v} \rangle_p$ and the corresponding $E_c$ for each system are shown in Fig. 1b. In both unstrained and strained cases, the WS$_2$ has the highest $\langle \mathbf{v} \rangle_p$ and the lowest $E_c$. Moreover, for all the materials, the tensile strain increases the $\langle \mathbf{v} \rangle_p$. These observations will be explained later. After reaching the $\langle \mathbf{v} \rangle_p$, the change of $\langle \mathbf{v} \rangle$ with $E$ is negligible for us-WSe$_2$ and us-MoSe$_2$, while other systems exhibit NDM to different degrees. The NDM can be characterized by the peak-to-valley ratio (PVR), which is ratio of the velocity at the peak to that at the valley of the NDM region. The PVRs for all systems are shown in Fig. 1c. In both unstrained and strained cases, the WS$_2$ has the largest PVR. Moreover, for all the materials, the tensile strain increases the PVR. These observations will be explained later. Based on these results, we conclude that the WS$_2$ is the best candidate for high-field applications, with the highest $\langle \mathbf{v} \rangle_p$, lowest $E_c$, and largest PVR; moreover, the tensile strain can generally improve the performance of MX$_2$.

To understand the trend of $\langle \mathbf{v} \rangle_p$ and $E_c$, we examine the relevant physical factors. As discussed in the Methods section, the electrons move in the momentum-space by $E$-driving drift and phonon-induced scattering. According to the Drude model, we have:

$$\langle \mathbf{v} \rangle \approx q \langle \tau \rangle \left\langle \frac{1}{m^*} \right\rangle \cdot \mathbf{E} , \qquad (3)$$

where $m^*$ is the "effective mass" tensor that depends on the electronic state and is calculated as: $\frac{1}{m^*} = \frac{1}{\hbar^2}\begin{pmatrix} \partial^2\varepsilon/\partial k_x\partial k_x & \partial^2\varepsilon/\partial k_x\partial k_y \\ \partial^2\varepsilon/\partial k_y\partial k_x & \partial^2\varepsilon/\partial k_y\partial k_y \end{pmatrix}$. Since here we choose the electric field to be along the zigzag (x) direction of MX$_2$, and record the velocity component along the x direction, thus the $1/m^*_{xx}$ is used. Note that the effective mass is a concept defined near band extrema. In our cases, we use it more loosely, basically as a state-dependent quantity. To remove the $E$ dependence from this equation, we consider the energy conservation requirement [38]:

$$q\mathbf{E}\cdot\langle\mathbf{v}\rangle \approx \langle\Delta S\rangle/\langle\tau\rangle, \tag{4}$$

where $\langle\Delta S\rangle$ is average energy difference between initial and final state for each scattering (termed average "scattering energy"). Combining Eq. 3 and 4, we finally get:

$$|\langle\mathbf{v}\rangle| \approx \sqrt{\langle\Delta S\rangle/\langle m^*\rangle}. \tag{5}$$

This equation can also be derived in other ways. [39,40] Fig. 2a compares the $\sqrt{\langle\Delta S\rangle/\langle m^*\rangle}$ with the magnitude of $\langle\mathbf{v}\rangle$ (for us-WS$_2$ as an example). Indeed, these two quantities agree reasonably well over a wide range of $E$. Eq. 5 indicates that the velocity is closely related with the $\langle m^*\rangle$ and the scattering energy: a larger scattering energy and a smaller $\langle m^*\rangle$ will give a higher velocity. Fig. 2b compares the $\langle\mathbf{v}\rangle_p$ with the $\sqrt{\langle\Delta S\rangle/\langle m^*\rangle}$ at $E_c$ across different materials. The strong correlation between these two quantities suggests that the $\sqrt{\langle\Delta S\rangle/\langle m^*\rangle}$ can be used as a "descriptor" to explain (and predict) the $\langle\mathbf{v}\rangle_p$: the reason why WS$_2$ has the highest $\langle\mathbf{v}\rangle_p$ is that it has a high $\sqrt{\langle\Delta S\rangle/\langle m^*\rangle}$ at $E_c$, and the strain increases the $\langle\mathbf{v}\rangle_p$ by increasing the $\sqrt{\langle\Delta S\rangle/\langle m^*\rangle}$. By further comparing the $\langle m^*\rangle$ and $\langle\Delta S\rangle$ across different systems, we find that the high $\sqrt{\langle\Delta S\rangle/\langle m^*\rangle}$ of WS$_2$ is mainly due to its smallest $\langle m^*\rangle$ (Fig. 2c), and the increase of $\sqrt{\langle\Delta S\rangle/\langle m^*\rangle}$ by tensile strain is mainly because of the increase of $\langle\Delta S\rangle$ (Fig. 2c).

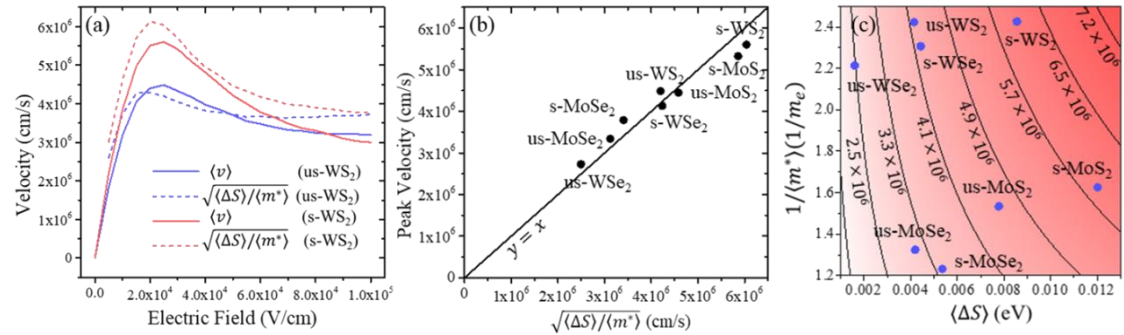

**Figure 2.** (a) Average velocity $\langle\mathbf{v}\rangle$ (solid lines) and $\sqrt{\langle\Delta S\rangle/\langle m^*\rangle}$ (dashed lines) for unstrained (us) and strained (s) WS$_2$, as a function of electric field. The $\langle\Delta S\rangle$ is the average scattering energy and the $\langle m^*\rangle$ is the average effective mass (see the main text for definition) (b) Comparison between peak velocity $\langle\mathbf{v}\rangle_p$ and $\sqrt{\langle\Delta S\rangle/\langle m^*\rangle}$ at $E_c$ for all the MX$_2$. (c) Distribution of $\langle\Delta S\rangle$ and $1/\langle m^*\rangle$; the curves connect the points that give the same $\sqrt{\langle\Delta S\rangle/\langle m^*\rangle}$ with the values labelled.

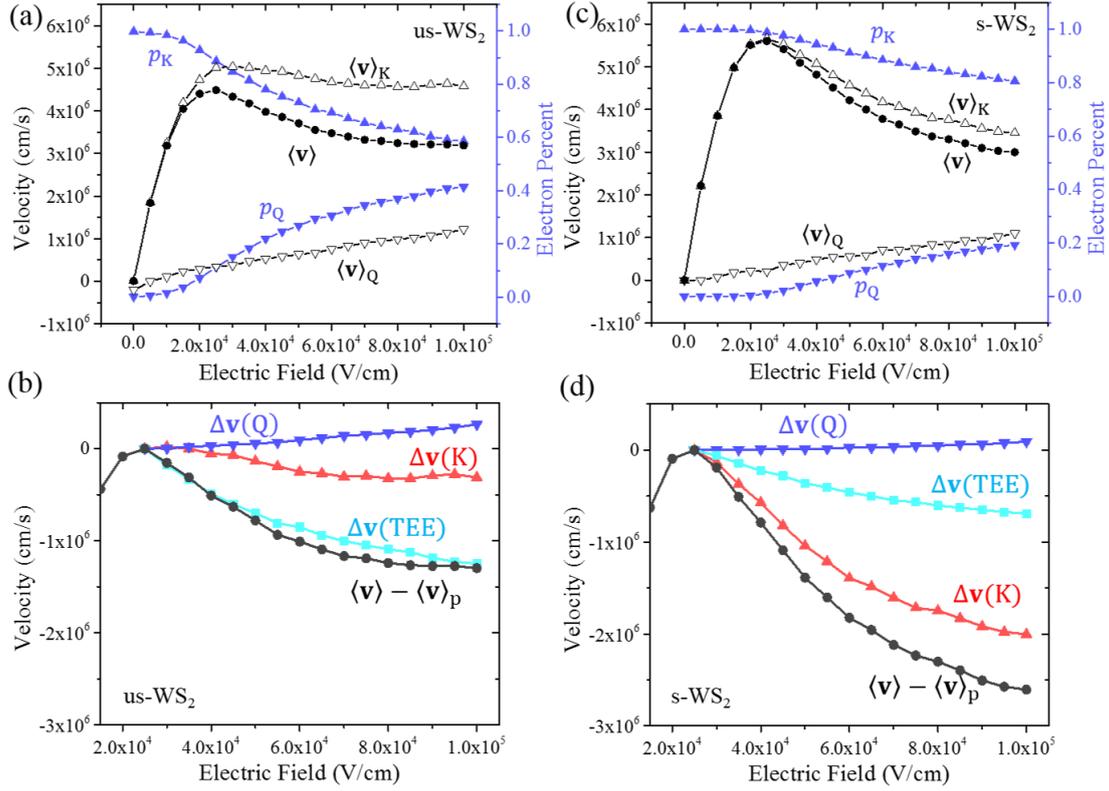

**Figure 3.** (a) Average velocity of electrons in K valley ($\langle v \rangle_K$), in Q valley ($\langle v \rangle_Q$), and in the system ($\langle v \rangle$), as well as the percent of electrons in K valley ($p_K$) and Q valley ($p_Q$), as a function of electric field, for us-WS$_2$. (b) Contributions of $\langle v \rangle_K$ change ($\Delta v(K)$), $\langle v \rangle_Q$ change ($\Delta v(Q)$), and the transferred electron effect ($\Delta v(TEE)$) to the $\langle v \rangle - \langle v \rangle_p$ in us-WS$_2$. See the text for how to quantify these contributions. (c) Similar to (a) but for s-WS$_2$. (d) Similar to (b) but for s-WS$_2$.

After understanding the peak velocity, we now focus on the NDM. When $E > E_c$, many MX$_2$ exhibit NDM except us-WSe$_2$ and us-MoSe$_2$. For all the MX$_2$ considered here, the CBM is located at the K point (K valley), and there is an additional valley that is located at the Q point (Q valley), which has a higher energy and larger effective mass. Therefore, the $\langle v \rangle$ can be expressed as:

$$\langle v \rangle = p_K \langle v \rangle_K + p_Q \langle v \rangle_Q, \qquad (6)$$

where $p_K$ ($p_Q$) is the percent of electrons in K (Q) valley ($p_K + p_Q = 1$), and $\langle v \rangle_K$ ($\langle v \rangle_Q$) is the average velocity of electrons in K (Q) valley, respectively. Increasing $E$ would increase the energy of electrons, decreasing (increasing) the $p_K$ ($p_Q$) and thus leading to electrons transfer from K to Q valley. Since the Q valley has a larger effective mass and hence $\langle v \rangle_Q < \langle v \rangle_K$, the electron transfer would decrease the $\langle v \rangle$. This effect is known as "transferred electron effect", TEE, which is the origin for the NDM of GaAs [41] at high field. Meanwhile, the $\langle v \rangle_Q$ and/or $\langle v \rangle_K$ can also change with $E$, which may act against the TEE. The competition of these three factors determines the magnitude of NDM. Fig. 3a shows how the valley-resolved populations and velocities change with $E$ for us-WS$_2$. Indeed, compared with the values at $E_c$, the $p_K$ ($p_Q$) decreases (increases), the $\langle v \rangle_Q$ also increases, while $\langle v \rangle_K$ decreases slightly. To quantify their contributions to the NDM, we take the differential of $\langle v \rangle$ with respect to $E$. Using Eq. 6 and $p_K + p_Q = 1$, the differential can be written as:

$$\frac{d\langle \mathbf{v} \rangle}{dE} = p_K \frac{d\langle \mathbf{v} \rangle_K}{dE} + p_Q \frac{d\langle \mathbf{v} \rangle_Q}{dE} + \frac{dp_Q}{dE}(\langle \mathbf{v} \rangle_Q - \langle \mathbf{v} \rangle_K). \tag{7}$$

The first term on the right side of the equation can be interpreted as the contribution of the $\langle \mathbf{v} \rangle_K$ differential to the $\langle \mathbf{v} \rangle$ differential (this contribution is proportional to the $p_K$). Similarly, the second term can be regarded as the contribution of the $\langle \mathbf{v} \rangle_Q$ differential to the $\langle \mathbf{v} \rangle$ differential, and the last term represents the contribution of the population differential. Integrating each term of Eq. 7 from $E_c$ to a higher filed $E_h$, the difference in velocity between $E_h$ and $E_c$ can be expressed as:

$$\begin{aligned}
\langle \mathbf{v} \rangle - \langle \mathbf{v} \rangle_p &= \int_{E_c}^{E_h} p_K \frac{d\langle \mathbf{v} \rangle_K}{dE} dE + \int_{E_c}^{E_h} p_Q \frac{d\langle \mathbf{v} \rangle_Q}{dE} dE + \int_{E_c}^{E_h} \frac{dp_Q}{dE}(\langle \mathbf{v} \rangle_Q - \langle \mathbf{v} \rangle_K) dE \\
&= \Delta \mathbf{v}(K) + \Delta \mathbf{v}(Q) + \Delta \mathbf{v}(TEE).
\end{aligned} \tag{8}$$

where $\Delta \mathbf{v}(K)$, $\Delta \mathbf{v}(Q)$ and $\Delta \mathbf{v}(TEE)$ quantify the contributions of $\langle \mathbf{v} \rangle_K$ change, $\langle \mathbf{v} \rangle_Q$ change, and TEE to the difference between $\langle \mathbf{v} \rangle$ and $\langle \mathbf{v} \rangle_p$. The $\Delta \mathbf{v}(TEE)$ is always negative for the band structure with two valleys where the larger mass valley has the higher energy, and a more negative $\Delta \mathbf{v}(TEE)$ characterizes a stronger TEE. Fig. 3b shows the $\langle \mathbf{v} \rangle - \langle \mathbf{v} \rangle_p$, $\Delta \mathbf{v}(K)$, $\Delta \mathbf{v}(Q)$ and $\Delta \mathbf{v}(TEE)$ for us-WS$_2$. Clearly, the $\langle \mathbf{v} \rangle - \langle \mathbf{v} \rangle_p$ is dominated by the $\Delta \mathbf{v}(TEE)$, indicating that the TEE is the main reason for the NDM. Compared with other us-MX$_2$, the us-WS$_2$ has the most negative $\Delta \mathbf{v}(TEE)$, which is the main cause of its largest PVR (see Fig. S6 for the valley-resolved velocities and populations of other us-MX$_2$, and Fig. S7 for the corresponding $\Delta \mathbf{v}(K)$, $\Delta \mathbf{v}(Q)$ and $\Delta \mathbf{v}(TEE)$). Particularly, compared with us-MoS$_2$ and us-MoSe$_2$, us-WS$_2$ has more electrons transferred; while compared with us-WSe$_2$, the us-WS$_2$ has a larger velocity difference between valleys; according to the expression of $\Delta \mathbf{v}(TEE)$ in Eq. 8, the more electrons transferred, and/or the larger velocity difference between two valleys, the more negative $\Delta \mathbf{v}(TEE)$ will be. Therefore, the us-WS$_2$ shows the strongest TEE.

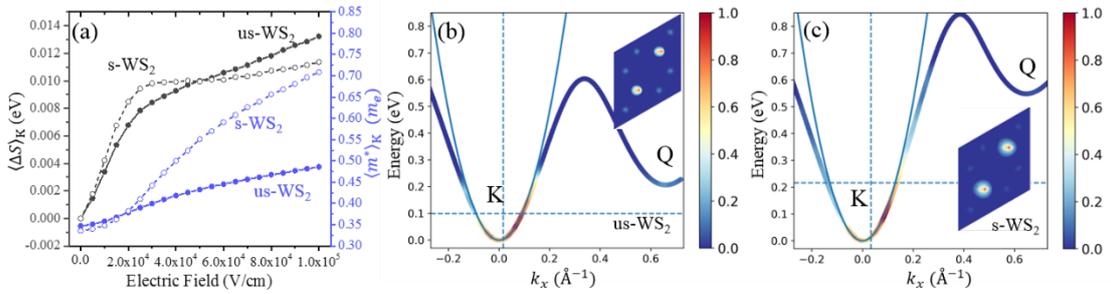

**Figure 4.** (a) Average scattering energy ($\langle \Delta S \rangle_K$) and average effective mass of electrons in K valley ($\langle m^* \rangle_K$) versus electric field for unstrained (us) and strained (s) WS$_2$. (b) Band structure and electron distribution of us-WS$_2$ at $4 \times 10^4$ V/cm electric field. The dashed lines show the average positions of the energy and the wavevector for electrons in K valley, and the thin solid line shows a parabola fitted to the CBM. The colors show the distribution of electrons. The inset shows the unit cell of the reciprocal lattice colored by electron distribution. (c) Similar to (b) but for s-WS$_2$.

Interestingly, we find that with 2% isotropic tensile strain, the main cause of the NDM in WS$_2$ is no longer TEE, instead, the decrease of $\langle \mathbf{v} \rangle_K$ with E (in other words, the NDM of electrons in K valley) is the main driving force for the overall NDM. This is evident in Fig. 3d, where the

⟨**v**⟩ − ⟨**v**⟩$_p$ is dominated by Δ**v**(K) rather than Δ**v**(TEE). Compared with the us-WS$_2$, the Δ**v**(TEE) becomes less negative while the Δ**v**(K) turns more negative, thereby changing the NDM mechanism.

Why does the tensile strain make the Δ**v**(K) (Δ**v**(TEE)) more (less) negative? The increase of Δ**v**(TEE) (i.e. the suppression of TEE) can be explained by the increase of ΔE$_{K-Q}$ (0.21 eV for us-WS2 and 0.55 eV for s-WS2), which becomes too large for effective electron transfer. This can be seen by comparing Fig. 3a with Fig. 3c: in us-WS$_2$, 30% electrons are transferred from K to Q valley when $E$ increases from $E_c$ to $10^5$ V/cm, while in s-WS$_2$, only 18% are transferred. The decrease of Δ**v**(K) is consistent with the more significant drop of ⟨**v**⟩$_K$ after $E_c$ (i.e. more pronounced NDM in K valley) in s-WS$_2$. To understand the origin of the strain-enhanced NDM in K valley, we plot the ⟨$m^*$⟩ and ⟨$ΔS$⟩ of electrons in K valley [38] as a function of $E$, as shown in Fig. 4a. Comparing these quantities between us-WS$_2$ and s-WS$_2$, we find that the ⟨$m^*$⟩$_K$ of s-WS$_2$ increases much more significantly after $E_c$ than that of us-WS$_2$, which is the main reason for their difference in K-valley NDM. The different behaviors of ⟨$m^*$⟩$_K$ can be attributed to different distributions of electrons, which are determined by the competing effects of the K and Q valleys. The $E$ drives the electrons to drift away from the CBM of the K valley, while the Q valley suppresses this displacement by greatly increasing the scatterings (Fig. S4). The strain increases the ΔE$_{K-Q}$ and thus the electrons are enabled to enter the regions further from the CBM (see Fig. 4b and c for the electron distributions). These regions are usually (more) non-parabolic and have larger $m^*$ compared with the CBM. Therefore, the ⟨$m^*$⟩$_K$ can increase over a larger range in strained material, making the Δ**v**(K) more negative and resulting in an enhanced NDM for the electrons in K valley.

Based on these understandings, one would expect that the further increase of ΔE$_{K-Q}$ would expose band regions with even larger ⟨$m^*$⟩$_K$ and thus further decreases the Δ**v**(K). Indeed, we find that a larger (3%) strain leads to a larger ΔE$_{K-Q}$ and a more negative Δ**v**(K) (Fig. S8). Consequently, the PVR increases to 1.95 (Fig. S8), larger than that of us-WS$_2$ (1.39) and 2% s-WS$_2$ (1.86). Moreover, we find that the 2% isotropic tensile strain also increases the NDM for other MX$_2$ (Fig. 1). Further analyses of their band structures and the valley-resolved properties show that for all the materials, the strain increases the valley separation while decreases the Δ**v**(K) (Fig. S7). The Δ**v**(K) decrease can thus be also related with the valley separation, which exposes the non-parabolic region of the K valley. Since it is a general phenomenon that the band gets more non-parabolic when being further from the valley edge, we anticipate that the valley-separation induced/enhanced NDM of the electrons in low-energy valley could be also observed in other materials beyond 2D MX$_2$ studied here.

**Conclusions:**
In summary, combining first-principles calculations and Monte Carlo simulations, we have studied the high field transport properties of electrons in unstrained and tensilely strained MoS$_2$, MoSe$_2$, WS$_2$ and WSe$_2$ monolayers. WS$_2$ is identified as the best candidate for high-field applications, with the highest peak velocity and the largest negative differential mobility, which require the lowest electric field. Tensile strain can improve these properties for all MX$_2$. The physical factors underlying these observations are uncovered: the peak velocity is determined by the ratio of scattering energy to effective mass; while the negative differential mobility can originate from either the electron transfer from K to Q valley (unstrained cases), or the non-parabolicity of the K valley (strained cases). Moreover, we find a general phenomenon that the valley separation can induce/enhance the negative differential mobility for electrons in low-energy valley, adding a new consideration for designing negative differential resistance devices.

**Acknowledgements:**

This work is supported by the Welch Foundation (Grant No. F-1959-20180324). This work used computational resources at (1) National Renewable Energy Lab (sponsored by the DOE's Office of EERE), (2) the Extreme Science and Engineering Discovery Environment (XSEDE) through allocation TG-CHE190065, (3) the Center for Nanoscale Materials (a DOE Office of Science user facility supported under Contract No. DE-AC02-06CH11357) at Argonne National Lab, and (4) the Center for Nanophase Materials Sciences (a DOE Office of Science user facility) at Oak Ridge National Lab.